\def\ltsima{$\; \buildrel < \over \sim \;$}    % Use in text mode
\def\simlt{\lower.5ex\hbox{\ltsima}}           % Use in math mode
\def\gtsima{$\; \buildrel > \over \sim \;$}    % Use in text mode
\def\simgt{\lower.5ex\hbox{\gtsima}}           % Use in math mode

\def\Bo{B_{\rm o}}
\def\Brms{B_{\rm rms}}
\def\EVS{Enrique V\'azquez-Semadeni}
\def\gameff{\gamma_{\rm eff}}
\def\o{{\rm o}}
\def\Peq{P_{\rm eq}}
\def\rhoc{\rho_{\rm c}}
\def\rhoic{\rho_{\rm IC}}
\def\Teq{T_{\rm eq}}
\def\u{{\bf u}}
\def\VS{V\'azquez-Semadeni}

\documentstyle[rmaaconf]{article}

\begin{document}

\title{MHD TURBULENCE, CLOUD FORMATION AND STAR FORMATION IN THE ISM}

\author{\EVS,}

\affil{Instituto de Astronom\'{\i}a, Universidad Nacional Aut\'onoma de M\'exico\\
Apdo. Postal 70-264, 04510 M\'{e}xico D.F., M\'{e}xico} 

\author{Thierry Passot and Annick Pouquet}

\affil{Observatoire de Nice, BP 229, 06340 Nice CEDEX 4, France}

\begin{resumen}
Discutimos el papel de la turbulencia en la formaci\'on de nubes y estrellas,
seg\'un se observa en simulaciones num\'ericas del medio interestelar. La
compresi\'on turbulenta en las interfases entre corrientes de gas en colisi\'on
es responsable de la formaci\'on de nubes de tama\~no intermedio ($\simlt 100$
pc) y peque\~nas (algunas decenas de pc), aunque las m\'as peque\~nas tambi\'en
se pueden formar por la fragmentaci\'on de c\'ascaras en expansi\'on alrededor
de centros de calentamiento por estrellas. Los m\'as grandes complejos de nubes
(varios cientos de pc) parecen formarse por un lento proceso de fusi\'on de
nubes individuales promovido por la inestabilidad gravitacional. Este proceso
se puede describir como una tendencia hacia la homogeneizaci\'on del medio a
gran escala debida a la gravedad, m\'as que como colisiones entre nubes. Estos
mecanismos operan tambi\'en en presencia del campo magn\'etico y de la 
rotaci\'on,
aunque con ligeras variaciones en la compresibilidad del flujo 
y la morfolog\'\i a de las
nubes que dependen de la intensidad y topolog\'\i a del campo. En resumen, el
papel de la turbulencia parece ser doble: los modos turbulentos 
peque\~nos contribuyen al soporte de las nubes contra su autogravedad, 
mientras que los modos grandes pueden tanto formar como destru\'\i r nubes.
\medskip
\end{resumen}

\begin{abstract}
We discuss the role of turbulence in cloud and star formation, as observed in
numerical simulations of the interstellar medium. Turbulent compression at the
interfaces of colliding gas streams is responsible for the formation
of intermediate ($\simlt 100$ pc) and small clouds (a few tens of pc), although
the smallest clouds can also form from fragmentation of expanding shells around
stellar heating centers. The largest cloud complexes (several hundred pc) seem
to form by slow, gravitational instability-driven merging 
of individual clouds, which can
actually be described as a large-scale tendency towards 
homogenization of the flow due to gravity rather than cloud collisions.
These mechanisms operate as well in the 
presence of a magnetic field and rotation, although 
slight variations on the compressibility and cloud morphology are present which
depend on the strength and topology of the field.
In summary, the role of turbulence in the life-cycle of clouds appears to be
twofold: small-scale modes contribute to cloud support, while large-scale modes
can both form or disrupt clouds.
\end{abstract}


\section{Introduction}
The interstellar medium (ISM) is a highly compressible turbulent flow
(e.g. Dickman 1985, Scalo 1987, Falgarone 1989) in which turbulent density
fluctuations are likely to be ubiquitous. With a few exceptions, 
however, most treatments of cloud
dynamics and formation including turbulence have traditionally considered only
the contribution of small-scale turbulent motions
for cloud support against gravitational collapse
(e.g. Chandrasekhar 1951; Shu et al.\ 1987; 
Bonazzola et al.\ 1987; L\'eorat et al.\ 1990;
Elmegreen 1991; \VS\ \& Gazol 1995).
In actuality, turbulent motions at the scale of clouds or larger may have both
cloud-forming and disrupting effects, which may be respectively associated with
the compressive and shearing modes of the turbulence. 
Although early contentions were that, being supersonic,
it should dissipate rapidly (e.g.\ Goldreich \& Kwan 1974), later studies
suggested that the energy sources present in the Galaxy (mainly stellar
activity and Galactic differential rotation) could be enough to replenish 
the turbulence (e.g., Fleck 1980; see the review by Dickman 1985). Stellar 
energy injection originates mostly from OB winds and supernova (SN) 
explosions, and the average rates for the Galaxy 
have been determined observationally by a number of authors (e.g., Abbott
1982; Van Buren 1989). 

Cloud and star formation (hereinafter CF and SF, respectively)
at the interfaces between 
colliding flow streams have been discussed analytically by 
Hunter \& Fleck (1982) and Elmegreen (1993).
Low-resolution hydrodynamical simulations of colliding flow streams were
performed by Hunter et al.\ (1986), and the effects of stellar forcing in the
large-scale flow
have been simulated by Bania \& Lyon (1980), Chiang \& Prendergast (1985), 
Chiang \& Bregman (1988) and Rosen et al.\ (1993). Fully turbulent regimes 
in the vertical direction in the Galactic disk have been explored by
Rosen et al.\ (1993) and by Rosen \& Bregman (1995). However, all of these
calculations have omitted self-gravity and magnetic fields, 
and have employed somewhat unrealistic
SF schemes, thus rendering it impossible to 
discuss the full energy budget and the life cycles of the clouds that form.

In this paper we discuss the interplay between turbulence and CF and SF
in recent numerical simulations of 
the interstellar medium (ISM) we have performed, which 
incorporate stellar and diffuse
heating with a more realistic SF scheme, 
parameterized cooling, self-gravity and large-scale shear (\VS\ et
al.\ 1995, hereafter Paper I), and magnetic fields and rotation (Passot et
al.\ 1995, hereafter Paper II).
In \S 2 we briefly describe the model system,
in \S\S 3 and 4 we discuss results without and with magnetic fields, and
in \S 5 we summarize the main conclusions.

\section{THE MODEL}
In the numerical calculations we consider a square region of the ISM 1 kpc
on a side on the Galactic plane. 
The hydrodynamic equations are solved in two dimensions
using a pseudospectral scheme with periodic boundary conditions. 
Model terms are included for the diffuse and stellar
heating and the parameterized cooling. Additionally, a large-scale sinusoidal
shear profile is imposed on the flow. In cases with rotation and
magnetic fields, the Coriolis and Lorentz forces are added to the momentum
equation, together with the induction equation. In all cases, 
Poisson's equation for the gravitational potential arising from 
the density {\it fluctuations} is also solved. Fiducial values of all
parameters are used that correspond to realistic values for the ISM, within the
uncertainties reported in the literature. Details on the numerical
scheme can be found in Papers I and II.

All variables are normalized to values which can be
considered typical of the ISM at the scale of 1 kpc. Thus, the units of
density, temperature and velocity are respectively $\rho_\o=1$ cm$^{-3}$, 
$T_\o = 10^4$ K and $u_\o = 11.7$ km s$^{-1}$. Note that $u_\o$ is chosen 
equal to the adiabatic speed of sound at $T_\o$.

Stellar heating 
is modeled as a local heating center a few pixels across that is turned 
on when the local value of the density exceeds a threshold $\rhoc$ and 
locally $\nabla \cdot \u < 0$, and is turned off after a time equal to 
the typical lifetime of OB stars ($\sim 5 \times 10^6$ yr). This mimics
the heating from ionizing radiation from massive stars. The diffuse heating
is taken to be proportional to $\rho^{-\alpha}$, with $0<\alpha<1$, 
mimicking the 
self-shielding of clouds against background UV radiation. Finally, radiative
cooling is parameterized as a term $\rho \Lambda$, where $\Lambda$ is a
piecewise power law of the temperature with exponent $\beta_i$ in the $i$-th
temperature interval, of the kind introduced by Chiang \& Bregman (1988).

In magnetic runs, the total magnetic field is split into a constant, uniform
azimuthal component ${\bf B}_\o$ and a fluctuating component ${\bf B}_{\rm
rms}$.

\section{NON-MAGNETIC RESULTS}
One of the most immediate results from the simulations using the fiducial 
parameter values is that the thermal (heating and cooling) time scales are 
much shorter than the dynamical (eddy turnover) time scales, a result already
pointed out by Elmegreen (1993). This has the
interesting consequence that diffuse heating and radiative cooling are almost
always in equilibrium with each other (except at sites of SF, where the 
dominant form of heating is stellar), and thus the temperature becomes 
exclusively a function of the density. In turn, the ideal-gas equation of 
state determines the thermal pressure, which then becomes a function of the
density as well, exhibiting a nearly polytropic behavior. Indeed, it was shown
in Papers I and II that, far from SF sites, the temperature and thermal
pressure are given by
$$\Teq = \Bigl[{\Gamma_\o \rhoic^\alpha \over \Lambda_i \rho^{1 +
\alpha}}\Bigr]^{1/\beta_i}\ \ \ \ \Peq={\rho \Teq \over \gamma} =
{\rho^{\gameff}\over \gamma}
\Bigl[{\Gamma_\o \rhoic^\alpha \over \Lambda_i}\Bigr]
^{1/\beta_i}$$
where $\gamma=5/3$ is the ratio of specific heats for the gas, 
$\rhoic \sim 0.2$ is a typical value of the density in the intercloud
medium (ICM), $\Gamma_\o$ is the diffuse heating rate at $\rho=\rhoic$,
$\Lambda_i$ is the coefficient of the cooling function in the
$i$-th temperature interval, and $\gameff= 1-{(1+\alpha) / \beta_i}$ is
an effective polytropic exponent. For the chosen values of $\alpha$ and
$\beta_i$, $\gameff$ is smaller than 0.5 below $T=10^5$ K. This implies that the
flow is highly compressible and, for practical purposes, the temperature 
and thermal pressure
can thus be taken as being {\it slaved} to the density field, which is
in turn controlled by the velocity field. 

\begin{figure}
%\vskip 2.5in
\def\epsfsize#1#2{0.6\hsize}
%\centerline{\epsffile{denvel62b.ps}}
\caption{
Contour plot of the density field with superimposed vector
velocity field for a $512 \times 512$ non-magnetic
run, at $t= 8 \times 10^7$ yr into the evolution. Note
that clouds are mostly located at the interfaces between colliding
flow streams.
}
\label{dv62b}
\end{figure}

In the simulations, expanding bubbles around SF sites (``HII regions'') are 
the only very-high-pressure regions and do not follow the above equilibrium
relations. However, by the time their heating ``stars'' turn off, they have
reached sizes of only a few tens of pc, and the shells they produce continue
to move inertially and rapidly merge with the general turbulent flow. 
Therefore, although the smallest clouds form from the disrupted shells, larger 
clouds basically constitute the turbulent density fluctuations in the flow,
arising mostly where larger-scale gas streams collide (fig.\ \ref{dv62b}). 
This result is in agreement with earlier speculations by 
Hunter \& Fleck (1982) and Elmegreen (1993).

\begin{figure}
%\vskip 2.5in
\def\epsfsize#1#2{0.3\hsize}
%\centerline{\epsffile{den30116.ps}}
\caption{
Contour plots of the density field for the same run as in fig.\ 1
%fig.\ \ref{dv62b}
at a) $t= 3.9 \times 10^7$  and b) $t= 15.1 \times 10^7$, showing the trend
towards homogenization at late times, apparently induced by gravity.
}
\label{den30116}
\end{figure}

The largest clouds in the simulations (several hundred pc) appear to form
by a combined effect of turbulence and gravity. Intermediate-sized clouds
appear to merge to form the largest complexes. However, these ``mergers'' 
do not constitute ``cloud collisions'' in the usual sense, as the process
is rather smooth and only noticeable when comparing epochs separated by several
$10^7$ yr. In fact, this effect can be thought of more as a large-scale
homogenization of the flow by gravity, as exemplified in
figs.\ \ref{den30116}a and \ref{den30116}b 
(see also the video accompanying 
Paper I). This effect is not observed in runs with reduced or zero
gravity, in which in fact the star formation rate (SFR) decays rapidly to zero.
The importance of gravity for large cloud formation is particularly interesting
considering that the initial choice of parameters ($\langle T \rangle = 10^4$
K, $\langle \rho \rangle = 1$ cm$^{-3}$) for the simulations is such
that the Jeans length is $L_{\rm J} = (\pi c^2/G\rho_\o)^{1/2} \sim 2$ kpc, or 
twice the size of the integration domain, at $t=0$. However, 
note that the initial
conditions are out of equilibrium and the system cools down to $\langle T
\rangle \sim 6,000$ K very rapidly, and that the turbulent velocity dispersion
in the domain is $\sim 6$ km s$^{-1}$. Furthermore, a gravitational instability
analysis considering the effective barotropic behavior given by eq.\ (1b) 
(Paper II) gives an effective Jeans length 
$$L_{\rm eff} = \Bigl[{\gameff \pi c^2 \over \gamma G\rho_\o} \Bigr]^{1/2} \sim
\hbox{ 0.8--1.1 kpc},$$
for $\gameff \sim 0.5$. Here, $c$ is taken as the sound speed corresponding to
the mean temperature in the simulations (including the ICM). Note that this
value of $c$ is typically slightly larger than the rms turbulent velocity
dispersion in the domain.
Thus, the domain can be effectively
gravitationally unstable after $t=0$. 
This issue was not explicitly stated in Paper I,
although it was speculated that self-gravity was responsible for the formation
of the largest clouds in the simulations.

Note that intermediate clouds that form from turbulent density 
fluctuations can also become
gravitationally unstable, as  suggested by the simulations in 
Paper II, which cover a range including lower values of $\gameff$. 
In fact,
simulations in which SF is not allowed (\VS\ \& Mendoza, 1995), 
exhibit collapse of intermediate-scale clouds generated by turbulent
fluctuations. Thus, SF itself prevents further generalized 
collapse of gravitationally
unstable clouds as it generates more turbulence (Franco \& Cox 1983). 
This mechanism may be at the
origin of the low efficiency of SF in the ISM (see, e.g., Evans 1991). Finally,
note that clouds in the simulations have strong velocity fluctuations within
them, due to either the turbulent velocities that formed them and/or the
turbulence generated by SF. Quiescent, bullet-like clouds
do not exist in our simulations, as also pointed out for real
clouds by Falgarone \& P\'erault (1988).

Concerning the efficiency of energy injection by the stars,
it was estimated in Paper I
that $\sim 0.05$ \% of the heating injected by stars ends up as turbulent
kinetic energy of the flow. Under these conditions, a self-sustaining cycle sets
in, in which turbulence induces SF through density fluctuations, and in
turn SF feeds the turbulence. Unfortunately, these results cannot be considered
conclusive evidence that stellar energy 
injection is enough to maintain the global
turbulence, since the stability of the cycle depends on the adjustable
parameter $\rhoc$. However, realistic massive-star formation 
rates are observed in the model ($\sim 10^{-4}$ OB stars kpc$^{-2}$ yr$^{-1}$), 
suggesting that the result is feasible.

In summary, three mechanisms of cloud formation appear to be dominant in
the simulations: intermediate-scale clouds form mainly by turbulent density
fluctuations at the interfaces of colliding gas streams; small-scale clouds
can form by either the same mechanism or from fragmentation of expanding
shells around SF sites. Finally, the largest cloud complexes form
through turbulent merging of intermediate clouds driven by gravitational
instability. Thus, the role of turbulence in the life cycle of clouds appears
to be twofold: small-scale (with respect to the cloud's size) turbulent
modes contribute towards cloud support, while large-scale modes provide either
cloud-forming or cloud-disrupting mechanisms.
However, an important caveat must be mentioned here: the simulations clearly do
not have turbulent modes at scales larger than the largest clouds. The effects
of these modes on the kpc-sized clouds is thus 
not represented, even though they could
drastically affect the conclusions based on small-scale turbulence. In that
context, simulations of the entire Galactic disk would be necessary.

\section{MAGNETIC EFFECTS}

\subsection{Effects of the magnetic field on cloud and star formation}

A linear instability analysis for the Galactic disk including an azimuthal
magnetic field has been carried out by Elmegreen (1991; 1994). The
corresponding analysis for the two-dimensional system of our simulations is
given in Paper II. In the absence of shear, a weak magnetic field
stabilizes
the medium by opposing collapse of radial perturbations, while a strong field
is destabilizing by preventing Coriolis spin-up of azimuthal perturbations
(magnetic braking). Since realistic perturbations should have both radial and
azimuthal components, the shear-less magnetic case is always unstable. 
In the presence of strong enough shear, the
field becomes stabilizing again, as azimuthal perturbations are sheared into
the radial direction before they have time to collapse.

\begin{figure}
%\vskip 2.5in
\def\epsfsize#1#2{0.3\hsize}
%\centerline{\epsffile{b0sf.ps}}
\caption{
Time integral of the star formation (SF) rate for various runs as
a function of the initial value of the uniform component of the
magnetic field $\Bo$. At small
$\Bo$, SF is inhibited because $\Bo$ is small enough not to counteract magnetic
braking, but is able to prevent radial collapse of sheared
condensations. Intermediate values of $\Bo$ counteract magnetic braking and
thus
promote SF. Very large values of $\Bo$ inhibit SF again because the magnetic
field rigidizes the medium.
}
\label{b0sf}
\end{figure}
 
In the turbulent regime, the cloud formation mechanisms described for the
non-magnetic cases continue to hold, although somewhat modified by the magnetic
field by criteria similar to those obtained from the instability analysis.
Indeed, in Paper II it was found that 
fully turbulent simulations exhibit varying degrees of
compressibility and SF as the initial uniform azimuthal 
component of the field $\Bo$ is increased.
This effect is shown in fig.\ \ref{b0sf}, 
which gives the time-integral of the SFR
for various runs with progressively higher values of $\Bo$. Moderate values are
seen to decrease the overall SFR, while larger values are seen to increase it.
Interestingly, saturation at very large $\Bo$ appears to occur, which can be
interpreted as a global rigidization of the medium by the field. In fact, ``HII
regions'' in the simulations do not expand nearly as much in the presence of
$\Bo$ as they do in non-magnetic cases. This ``pressure cooker''
effect is similar to the results
of Slavin \& Cox (1993) for the expansion of SN remnants in the presence of
magnetic fields. In fact, in the simulations, this cloud-confining effect of
the magnetic field is as notorious as its cloud-supporting effect.

\begin{figure}
%\vskip 2.5in
\def\epsfsize#1#2{0.25\hsize}
%\centerline{\epsffile{lessfilnolab.ps}}
\caption{
Contour plots of the density field at $t=5.2 \times 10^6 yr$
for three runs with
$\Bo=0, 1.5$ and $10 \mu$G (left to right), but otherwise identical. Note the
tendency towards more roundish structures as $\Bo$ increases.
}
\label{lessfil}
\end{figure}

An unexpected effect of the uniform field component is that clouds appear to be
slightly {\it more roundish, less filamentary} at larger values of $\Bo$
(fig.\ \ref{lessfil}),
contrary to the common belief that the opposite effect should occur. We
speculate that this phenomenon is due to a combination of factors: first, 
the above mentioned ``pressure cooker'' effect which does not allow shells to
expand too much; second, gas motions along field lines, when promoted by
compressions (as opposed to body forces like gravity), cause an increase in the
field component perpendicular to the motion, opposing it.

The effects of a uniform azimuthal field are to be contrasted to those of an
initial small-scale fluctuating field $\Brms$. In this case, magnetic tension
in the field lines induces small scale compressions in the flow, increasing the
global compressibility (fig.\ 5a) of the medium.

\subsection{Effects of star formation on the magnetic field}

\begin{figure}
%\vskip 2.5in
\def\epsfsize#1#2{0.25\hsize}
%\centerline{\epsffile{fig5ab.ps}}
\caption{
a) (Left) evolution of the ratio of compressible to total kinetic energy
for two runs with different initial values of the fluctuating component of the
field. The characteristic scale of the initial fluctuations is 1/4
the size of the integration box. Initially the run with stronger
fluctuations has larger compressibility. Due to the inverse magnetic cascade
and stellar activity,
however, by $t \sim 1.5 \times 10^8$ yr the fluctuating magnetic energies
of the two runs are comparable, causing also comparable
compressibilities. b) (Right) evolution of the fluctuating component of
the magnetic energy for three runs with different initial values of the uniform
component of the field.  Both the amplitude of the fluctuations and their time
derivative are seen to increase with $\Bo$.
}
\label{fig5ab}
\end{figure}
 
Stellar heating being the ultimate source of energy for the turbulence in 
the simulations, it has important effects on the magnitude and topology of the
magnetic field as well. One of the most important results is that SF appears to
be capable of maintaining and even increasing the total magnetic energy in the
flow by means of the compressible, non-periodic forcing it exerts. 
This result is all the more interesting considering that the simulations are
two-dimensional, and therefore cannot support dynamo activity.

The mechanism apparently responsible for the magnetic amplification in our
simulations requires the presence of the uniform component of the field
$\Bo$ (Paper II). This is exemplified in fig.\ 5b, which shows the evolution of
the fluctuating magnetic energy for three runs with different values of $\Bo$.
Clearly, both the amplitude of the fluctuations and their time derivative are
larger at larger $\Bo$.

\begin{figure}
%\vskip 2.5in
\def\epsfsize#1#2{0.6\hsize}
%\centerline{\epsffile{denmag32.ps}}
\caption{
Density and magnetic fields for a $512 \times 512$
run at $t=.42 \times 10^8$. Note the
strong magnetic turbulence inside clouds and the rather smooth character
of the magnetic field in the ICM. Regions of alignment of the magnetic field
and density features can be seen, for example, 
in the filament in the upper right corner.
However there are also regions where the magnetic field is perpendicular
to the density features such as the lower portion of the same cloud
and also the cloud near the center of the lower left quadrant.
}
\label{denmag32}
\end{figure}

An interesting feature of the dynamics in the simulations is that there seems
to be no single predominant physical agent. As much as the field constrains
fluid and cloud motions to some extent, 
the field is in turn strongly distorted by
the turbulence and SF. In particular, in clouds formed by turbulent
compressions, the field exhibits a certain tendency towards alignment with
elongated density features, as the field component perpendicular to the
compression is amplified by flux freezing. This mechanism has the further
consequence that the field is larger in general inside clouds than in the ICM
by factors of $\sim 4$ ($\sim 3 \mu$G for the ICM and $\sim 12 \mu$G within
clouds), with excursions up to factors $\sim 10$. Also, the strong turbulence
within the clouds causes the field to have strong fluctuations inside them 
(fig.\ 6).

It is noteworthy that the stellar energy injection occurs at scales small 
compared with the size of the system, and thus kinetic and magnetic
energy transfer to larger scales is observed. Although observed here in a
2D context, it is known that in 3D an inverse cascade of magnetic helicity
exists (as opposed to square magnetic potential in 2D),
with the magnetic energy following to a
lesser extent (Pouquet, Frisch \& L\'eorat, 1976; Meneguzzi,
Frisch \& Pouquet, 1981; Horiuchi \& Sato, 1986, 1988). It will be of great
interest to determine whether the observed cascades are still maintained in 3D
simulations with forcing at the small scales.

\section{CONCLUSIONS}

Turbulence is a major cloud-forming agent 
in numerical simulations of the ISM at the kpc scale 
(Papers I and II). The non-magnetic
simulations (Paper I) suggest that intermediate ($\simlt 100$ pc) and 
small (a few tens of pc)
clouds mainly form at the interfaces of colliding flow streams. The smallest 
clouds can
additionally form from fragmentation of expanding shells around stellar heating
centers. Large cloud complexes (several hundred pc) appear to form from
gravitational instability. However, since the medium is turbulent and contains
already sizable clouds, the formation of the large complexes proceeds through
gradual merging of the clouds that is almost imperceptible as one watches the
simulation evolve, but is noticeable when comparing epochs separated by
several $10^7$ yr. 

SF injects energy to the turbulence, and the simulations indicate that
a self-sustaining cycle may be established, although the caveat exists that the
simulations contain an adjustable parameter, namely the threshold density for
SF $\rhoc$, which controls the stability of the cycle.

The magnetic field introduces quantitative variations in these mechanisms, but
does not seem to alter them essentially, although the role of the Parker
instability cannot be evaluated by the simulations, as they neglect the
vertical direction in the Galaxy. The field modulates the density
contrast that can be achieved by turbulent compressions, as well as 
the morphology of the clouds. In turn, the turbulent magnetic energy is also fed
by stellar activity, provided a sizable uniform component of the field is
present. The stellar energy injection occurs at comparatively small scales, and
energy cascades to larger scales. Whether this is an artifact of the
two-dimensionality of the simulations remains to be evaluated by resorting
to three-dimensional computations.

In summary, the role of turbulence in the life-cycle of clouds appears to be
twofold: small-scale modes contribute to cloud support, while large-scale modes
can both form or disrupt clouds.

Finally, it should be pointed out that the simulations so far have not included
the energy injection from SNe and OB winds, which can inject energy at
significantly larger rates than heating from ionizing radiation from main
sequence OB stars. The largest turbulent 
velocity dispersion that can be imparted to the flow by stellar heating
is comparable to the thermal
velocity dispersion at the temperature of our ``HII regions'', $\sim 10^4$ K,
which roughly balances gravity for these complexes. 
In the non-magnetic case, this
turbulent energy is enough to blow the complexes apart, but not in the magnetic
case due to the ``pressure cooker'' effect.
The inclusion of SN/wind
energy should cause these clouds to be easily disrupted. Work in this direction
is in progress (Gazol et al.\ 1995).


\begin{references}

\reference
Abbott, D. C. 1982, ApJ 263, 723

\reference
Bania, T. M. \& Lyon, J. G. 1980, ApJ, 239, 173

\reference
Bonazzola, S., Falgarone, E., Heyvaerts, J., P\'erault, M., Puget, J. L.
1987, A\&A 172, 293

\reference
Chandrasekhar, S. 1951, Proc. R. Soc. London, 210, 26

\reference
Chiang, W.-H., \& Bregman, J. N. 1988, ApJ, 328, 427

\reference
Chiang, W.-H., \& Prendergast, K. H. 1985, ApJ, 297, 507

\reference
Dickman, R. L. 1985, in  Protostars and Planets II, ed.\ D. C. Black \&
M. S. Matthews (Tucson: Univ. of Arizona Press), 150

\reference
Elmegreen, B. G. 1991, ApJ, 378, 139

\reference
Elmegreen, B. G. 1993, ApJ, 419, L29

\reference
Elmegreen, B. G. 1994, ApJ, 433, 39
 
\reference
Evans, N. J., 1991, in Frontiers in Stellar Evolution, ed. D. L.
Lambert (San Francisco: A. S. P.), 45

\reference
Falgarone, E. 1989, in Structure and Dynamics of the Interstellar
Medium, IAU Colloquium 120, ed. G. Tenorio-Tagle, M. Moles \& J.
Melnick (Berlin: Springer), 68

\reference
Falgarone, E., \& P\'erault, M. 1988, AA, 205, L1

\reference
Fleck, R. C., Jr. 1980, ApJ 242, 1019

\reference 
Franco, J. \& Cox, D.P. 1983, ApJ, 273, 243.

\reference
Gazol, A., Passot, T., Pouquet, A. \& \VS, E. 1995, in preparation

\reference
Goldreich, P, \& Kwan, J. 1974, ApJ, 189, 441

\reference
Horiuchi, R. \& Sato, T. 1986, Phys. Fluids 26, 1161
 
\reference
Horiuchi, R. \& Sato, T. 1988, Phys. Fluids 31, 1142

\reference
Hunter, J. H., Jr., \& Fleck, R. C. 1982, ApJ, 256, 505

\reference
Hunter, J. H. Jr., Sandford, M. T. II, Whitaker, R. W. \& Klein, R. I.
1986, ApJ, 305, 309

\reference
L\'eorat, J., Passot, T., \& Pouquet, A. 1990, MNRAS, 243, 293

\reference
Meneguzzi, M., Frisch, U., \& Pouquet, A. 1981, Phys. Rev. Lett. 47, 1060

\reference
Passot, T., \VS, E. \& Pouquet, A. 1995, ApJ, 455, 536 (Paper II)

\reference
Pouquet, A., Frisch, U., \& L\'eorat, J. 1976, J. Fluid Mech., 77, 321

\reference
Rosen, A., Bregman, J. N., \& Norman, M. L. 1993, ApJ, 413, 137

\reference
Rosen, A. \& Bregman, J. N. 1995, ApJ, 440, 634

\reference
Scalo, J. M. 1987, in Interstellar Processes, ed. D. J. Hollenbach \&
H. A. Thronson (Dordrecht: Reidel), 349

\reference
Shu, F. N., Adams, F. C., Lizano, S. 1987, ARA\&A, 25, 23

\reference
Slavin, J. \& Cox, D. 1993, ApJ, 417, 187

\reference
Van Buren, D. 1989, ApJ, 338, 147

\reference
\VS, E., \& Gazol, A. 1995, A\&A, 303, 204

\reference
V\'azquez-Semadeni, E., Passot, T. \& Pouquet, A. 1995, ApJ, 441, 702 (Paper
I)

\reference
\VS, E. \& Mendoza, V. 1995, in preparation


\end{references}
\end{document}